\documentclass[journal]{IEEEtran}
\usepackage[linesnumbered,ruled]{algorithm2e}
\usepackage{graphicx,amssymb,mathrsfs,amsmath,array,color,amsthm}
\usepackage{subfigure}
\usepackage{multirow}
\usepackage{booktabs}
\usepackage{mathtools}
\usepackage{bm}
\usepackage{cite}
\usepackage[colorlinks,urlcolor=blue,driverfallback=dvipdfm]{hyperref}

\usepackage{pifont}
\newcommand{\cmark}{\ding{51}}
\newcommand{\xmark}{\ding{55}}

\newcommand{\norm}[1]{\lVert#1\rVert}

\begin{document}
\title{Decoupled-and-Coupled Networks: Self-Supervised Hyperspectral Image Super-Resolution with Subpixel Fusion}

\author{Danfeng Hong,~\IEEEmembership{Senior Member,~IEEE,}
        Jing Yao,
        Deyu Meng,~\IEEEmembership{Member,~IEEE,}
        Naoto Yokoya,~\IEEEmembership{Member,~IEEE,}
        and Jocelyn Chanussot,~\IEEEmembership{Fellow,~IEEE}

\thanks{This work was supported by the MIAI @ Grenoble Alpes (ANR-19-P3IA-0003), and the AXA Research.}
\thanks{D. Hong and J. Yao are with the Key Laboratory of Computational Optical Imaging Technology, Aerospace Information Research Institute, Chinese Academy of Sciences, Beijing 100094, China (e-mail: hongdf@aircas.ac.cn; yaojing@aircas.ac.cn).}
\thanks{D. Meng is with the School of Mathematics and Statistics, Xi’an Jiaotong University, 710049 Xi’an, China. (e-mail: dymeng@mail.xjtu.edu.cn)}
\thanks{N. Yokoya is with Graduate School of Frontier Sciences, the University of Tokyo, 277-8561 Chiba, Japan, and also with the Geoinformatics Unit, RIKEN Center for Advanced Intelligence Project (AIP), RIKEN, 103-0027 Tokyo, Japan. (e-mail: naoto.yokoya@riken.jp)}
\thanks{J. Chanussot is with the Univ. Grenoble Alpes, CNRS, Grenoble INP, GIPSA-Lab, 38000 Grenoble, France (e-mail: jocelyn@hi.is).}
}


\maketitle
\begin{abstract}
Enormous efforts have been recently made to super-resolve hyperspectral (HS) images with the aid of high spatial resolution multispectral (MS) images. Most prior works usually perform the fusion task by means of multifarious pixel-level priors. Yet the intrinsic effects of a large distribution gap between HS-MS data due to differences in the spatial and spectral resolution are less investigated. The gap might be caused by unknown sensor-specific properties or highly-mixed spectral information within one pixel (due to low spatial resolution). To this end, we propose a subpixel-level HS super-resolution framework by devising a novel decoupled-and-coupled network, called DC-Net, to progressively fuse HS-MS information from the pixel- to subpixel-level, from the image- to feature-level. As the name suggests, DC-Net first decouples the input into common (or cross-sensor) and sensor-specific components to eliminate the gap between HS-MS images before further fusion, and then fully blends them by a model-guided coupled spectral unmixing (CSU) net. More significantly, we append a self-supervised learning module behind the CSU net by guaranteeing the material consistency to enhance the detailed appearances of the restored HS product. Extensive experimental results show the superiority of our method both visually and quantitatively and achieve a significant improvement in comparison with the state-of-the-arts. Furthermore, the codes and datasets will be available at \url{https://sites.google.com/view/danfeng-hong} for the sake of reproducibility.
\end{abstract}
\graphicspath{{figures/}}


\section{Introduction}
\IEEEPARstart{R}{cent} Hyperspectral (HS) images are characterized by abundant and detailed spectral information, which enables the recognition of materials at a more accurate level compared to RGB or multispectral (MS) images. In recent years, HS imaging has been garnering increasing attention in a wide range of applications related to computer vision, such as image restoration \cite{arad2016sparse,he2021fast,sarkar2021non}, image super-resolution \cite{yang2017pannet,zhang2018exploiting,xu2019nonlocal,xue2021spatial,dong2021model}, object detection and tracking \cite{chakrabarti2011statistics,nasrabadi2013hyperspectral,tochon2017object,wu2019orsim,xiong2020material}, to name a few.

\begin{figure}[!t]
	  \centering
		\includegraphics[width=0.46\textwidth]{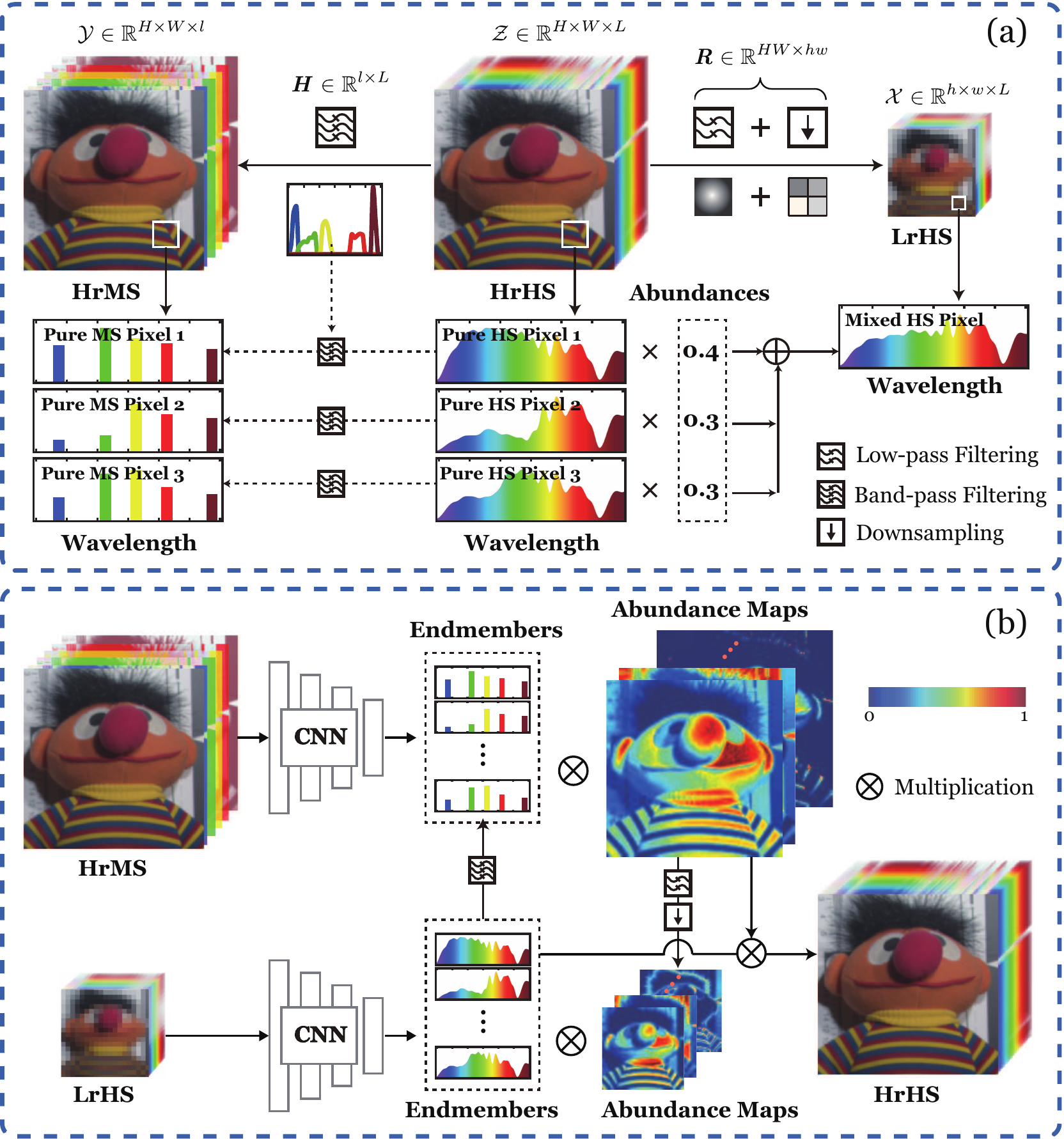}
        \caption{(a) shows the observation models for HrMS and LrHS images and illustrates potential challenges (highly mixed spectral signatures) in HS-SR. (b) provides a solution by coupled spectral unmixing (CSU) to decompose mixed pixels to pure and informative subpixels, yielding a subpixel-level HS-SR approach.}
\label{fig:motivation}
\end{figure}

HS images are capable of capturing more subtle discrepancies between different objects, which, to a great extent, is benefited from its high spectral resolution. As a trade-off, HS imaging systems are usually designed to acquire the data at a high spatial sampling distance (i.e., sacrificing spatial resolution), thereby limiting the range of potential applications in practice \cite{kawakami2011high}. But fortunately, MS imaging systems, with the broader spectral bandwidth, can provide finer spatial information. Therefore, fusing the low-resolution HS (LrHS) and high-resolution MS (HrMS) image pair is an intuitive and feasible solution to generate the high-resolution HS (HrHS) product, also known as MS-guided HS super-resolution (HS-SR). Fig.~\ref{fig:motivation} (a) illustrates the observation models for HrMS and LrHS images degraded from the HrHS image.

Over the last decade, a variety of model-driven HS-SR algorithms have been successfully developed \cite{yokoya2017hyperspectral}. These approaches model the underlying relationships between HS-MS data by various hand-crafted priors and the knowledge of relevant sensor characteristics, e.g., spectral response function (SRF) and point spread function (PSF) of different camera systems. Very recently, the great success of convolutional neural networks (CNNs) has also made significant progress on the HS-SR task \cite{dian2018deep,qu2018unsupervised,xie2019multispectral,yao2020cross,uezato2020guided,zhang2020deep}. Owing to the powerful representation ability, CNNs-based models can excavate the intrinsic properties that lie in the HS (or MS) image more effectively and blend HS and MS images in a more compact way. Nevertheless, those existing, either model-based or data-driven (e.g., CNNs), fusion approaches rarely consider underlying problems that impact the HS-RS performance to a great extent, e.g., different sensor-specific properties leading to the weak affinity between original HS and MS data, highly-mixed pixels widely existed in both HS and MS images (see Fig.~\ref{fig:motivation} (a)).

To this end, we propose a decoupled-and-coupled network (DC-Net) for the HS-SR task. DC-Net is a novel progressive fusion framework from the pixel- to subpixel-level fusion, from the image- to feature-level fusion. As the name suggests, we first design a decoupled subnetwork (D-Net) by means of generative adversarial networks (GANs) \cite{goodfellow2014generative}. The D-Net disentangles the input HS and MS images to common (or cross-sensor) and sensor-specific components and recombines them in a new image space, where the sensor-specific information is preliminary exchanged to enable a more tight connection of HS-MS information, further providing a better starting point for the subsequent fusion process. Inspired by spectral unmixing \cite{keshava2002spectral}, we then develop a model-driven coupled subnetwork (C-Net) on the output of the D-Net to address the mixed pixel problem for a more compact fusion of HS-MS images. Fig.~\ref{fig:motivation} (b) illustrates a subpixel-level solution for HS-SR by the CSU net. More significantly, a self-supervised learning module (S-Net) is deployed to the C-Net to further enhance detailed geometries and textures, yielding more realistic HrHS images. The main contributions of this paper are three-fold as follows.
\begin{itemize}
    \item Beyond the pixel-level fusion, we provide a new subpixel insight for the HS-SR task by developing a unified decoupled-and-coupled network architecture in an end-to-end fashion, DC-Net for short.
    \item DC-Net is a progressive fusion framework, which not only can eliminate the gap between HS-MS images for better pixel-level information exchange (from the image perspective via D-Net) but also can alleviate the effects of highly mixed spectral pixels (due to low spatial resolution) for a more compact subpixel-level fusion (from the feature perspective via C-Net).
    \item A plug-and-play self-supervised module, i.e., S-Net, is devised in C-Net by aligning the abundance maps obtained by unmixing HS and MS images, respectively, to maintain \textit{semantic consistency} between abundance maps of HS-MS images, further improving the quality of the HrHS product.
\end{itemize}

The rest of this paper is organized as follows. Section 2 introduces the related works from the perspectives of traditional models and deep models in detail. We then elaborate the proposed DC-Net and details each newly-developed modules. Experiential results are conducted in comparison with currently existing state-of-the-art HS-SR approaches in Section 4. Finally, Section 5 draws a conclusion with a possible future outlook.

\section{Related Work}
\label{sec:1}
\subsection{Traditional Models}
HS pansharpening is a heuristic way to perform the HS-MS fusion \cite{loncan2015hyperspectral}, which has been widely applied in the HS image super-resolution task. Component substitution (CS) and multiresolution analysis (MRA) are the two main types of pansharpening techniques. The former one aims to inject detailed information of MS images into the LrHS image, thereby generating the HrHS product. The latter one is to pansharpen the HS image by linearly combining MS bands to synthesize a HrHS band using regression techniques.

One representative technique that fuses the HS-MS images is the subspace-based approach, which are usually designed to enhance the spatial resolution of HS images by means of matrix factorization \cite{kawakami2011high,yokoya2011coupled,akhtar2014sparse,lanaras2015hyperspectral,dong2016hyperspectral} or Bayesian estimation \cite{akhtar2015bayesian,akhtar2016hierarchical}. Kawakami \textit{et al.} \cite{kawakami2011high} first decomposed the LrHS image into a spectral dictionary and corresponding sparse representations via matrix factorization, and the HrHS image can be then formed by using shareable sparse coefficients estimated from the RGB observation with a sampled spectral dictionary. Akhtar \textit{et al.} \cite{akhtar2014sparse} jointly considered different physical properties of materials in the scene, e.g., sparsity, non-negativity, and spatial structure, to improve the quality of the HrHS image.  Dong \textit{et al.} \cite{dong2016hyperspectral} structurized the sparse coding model with the application to HS-SR. Yokoya's algorithm \cite{yokoya2011coupled} is designed to perform the HS-MS fusion via an effective couple matrix factorization approach, which can be well explained from the spectral unmixing perspective. Similarly, Lanaras \textit{et al.} \cite{lanaras2015hyperspectral} further extended the \cite{yokoya2011coupled} by jointly unmixing the input HS and MS data and fusing them in the latent subspace. In \cite{akhtar2015bayesian}, the spectral dictionary is learned by using a parameter-free Bayesian model. Using the learned dictionary, the HrHS image can be reconstructed with potential sparse assumptions. Further, the same investigators used the Gaussian process as the priors on the basis of the Bayesian model to super-resolve the HS image \cite{akhtar2016hierarchical}. Beyond matrix factorization, authors of \cite{dian2017hyperspectral} and \cite{zhang2018exploiting} have attempted to model the HS-SR issue in tensor form to achieve better structural representations.

\begin{figure*}[!t]
	  \centering
			\includegraphics[width=0.98\textwidth]{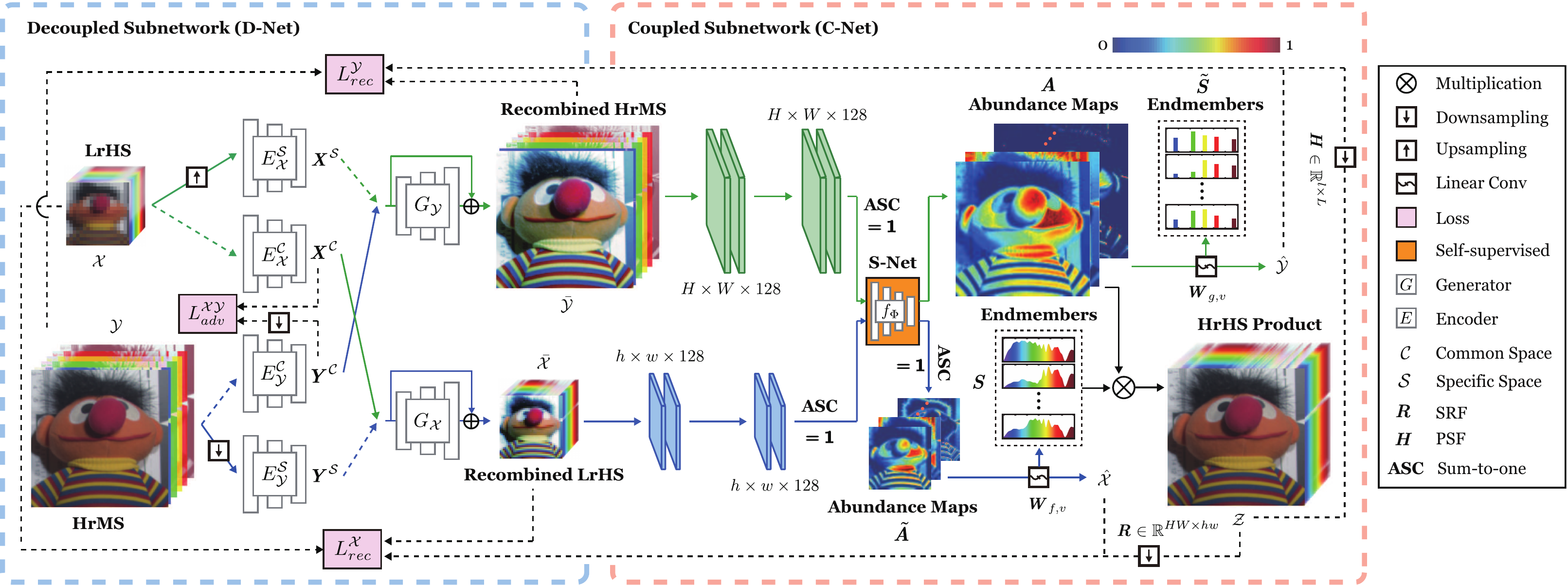}
        \caption{An illustrative overview of the proposed DC-Net. It consists of two subnetworks: D-Net and C-Net. With the adversarial loss $L_{adv}^{\mathcal{X}\mathcal{Y}}$ between the common content ($\bm{\bar{X}}$ and $\bm{\bar{Y}}$) of $\bm{X}$ and $\bm{Y}$, the reconstruction loss $L_{rec}$, and the self-supervised module (i.e., S-Net), we can learn an end-to-end HS-SR network more effectively without any prior training, and better enhance detailed appearances.}
\label{fig:workflow}
\vspace{-12pt}
\end{figure*}

\subsection{Deep Learning based Models}
These traditional methods have shown the competitive performance in the HS-MS fusion task, yet they, to a great extent, rely on cross-sensor calibration and strong hand-crafted priors. Inspired by the recent success of deep learning (DL) techniques, CNNs-based approaches have been garnering growing attention in the HS-SR task \cite{dian2018deep,qu2018unsupervised,xie2019multispectral,yao2020cross,uezato2020guided,zhu2020hyperspectral}. Dian \textit{et al.} \cite{dian2018deep} proposed a two-stage HS image pansharening framework by using CNNs-based prior training to refine the initialized fusion results obtained by traditional optimization methods. Qu \textit{et al.} \cite{qu2018unsupervised} trained an end-to-end unsupervised Dirichlet network to restore the HrHS image. However, the network needs to be alternatively optimized with given SRF and PSF of hybrid cameras. To maximize the HrHS image reconstruction accuracy, Fu \textit{et al.} \cite{fu2019hyperspectral} designed a CNNs-based HS-SR method by retrieving the optimal SRF from an external dataset. They further extended their work \cite{zhang2019hyperspectral} to select the best SRF by the joint use of external and internal learning for HS image reconstruction at a more accurate level. Xie \textit{et al.} \cite{xie2019multispectral} sought to open the ``black-box'' and presented an interpretable deep supervised model for the MS/HS fusion task. Yao \textit{et al.} \cite{yao2020cross} attempted to explain an unsupervised deep super-resolution network from the spectral unmixing perspective by introducing cycle consistency and cross-attention mechanism. Zhu \textit{et al.} \cite{zhu2020hyperspectral} proposed a supervised deep residual network in a progressive way for the HS image super-resolution.

\section{Decoupled-and-Coupled Networks}
\subsection{Overview}
A large distribution gap between HS-MS images possibly caused by unknown sensor-specific information (e.g., HS-MS) and highly mixed spectral pixels has been shown to be a challenging and potential problem in the HS-SR task \cite{yokoya2011coupled,lanaras2015hyperspectral,yokoya2017hyperspectral}. Focusing on it, we provide a point-to-point solution by designing a novel network architecture, i.e., DC-Net, which consists of two subnetworks. D-Net aims to decompose the original HS-MS data into the common content (or structure) and sensor-specific features, and implement crossover recombination on the sensor-specific information to improve the compatibility between HS-MS images for better subsequent fusion. With the input of recombined HS-MS images, C-Net performs the subpixel-level fusion by decomposing the mixed pixels into fractional coefficients (termed as \textit{abundances}) and pure spectral signatures (termed as \textit{endmembers}), yielding an interpretable CSU-guided HS-SR network. It should be noted that DC-Net (i.e., D-Net and C-Net) can be regarded as a novel progressive fusion framework from pixel- to subpixel-level, from the image- to feature-level. Further, a simple but effective self-supervised subnetwork (S-Net) in C-Net is devised to improve the quality of the HrHS product by correcting the material correspondence between HS-MS images. Fig.~\ref{fig:workflow} illustrates the architecture overview of the proposed DC-Net. 

\subsection{Problem Formulation}
Let $\mathcal{X}\in \mathbb{R}^{h\times w\times L}$, $\mathcal{Y}\in \mathbb{R}^{H\times W\times l}$, and $\mathcal{Z}\in \mathbb{R}^{H\times W\times L}$ be LrHS with $hw$ pixels by $L$ bands, HrMS with $HW$ pixels by $l$ bands, and HrHS images with $HW$ pixels by $L$ bands, respectively, ($h<H$, $w<W$, $l<L$), then $\bm{X}\in \mathbb{R}^{L\times hw}$, $\bm{Y}\in \mathbb{R}^{l\times HW}$, and $\bm{Z}\in \mathbb{R}^{L\times HW}$ are defined as the unfolded matrices. The observation models for LrHS and HrMS images from the HrHS image can be written as
\begin{equation}
\label{eq1}
\begin{aligned}
      &\bm{X} = \bm{Z}\bm{R} + \bm{N}_{x},\\
      &\bm{Y} = \bm{H}\bm{Z} + \bm{N}_{y},
\end{aligned}
\end{equation}
where $\bm{H}\in \mathbb{R}^{l\times L}$ and $\bm{R}\in \mathbb{R}^{HW\times hw}$ denote the SRF and PSF to degrade the spectral and spatial resolutions of $\bm{Z}$, respectively, and $\bm{N}_{x}$ and $\bm{N}_{y}$ are the observed noises.

Ideally, the observations in Eq. (\ref{eq1}) can be further unfolded by using spectral mixing model as
\begin{equation}
\label{eq2}
\begin{aligned}
      &\bm{X} = \bm{S}\bm{A}\bm{R} + \bm{N}_{x} = \bm{S}\bm{\tilde{A}} + \bm{N}_{x},\\
      &\bm{Y} = \bm{H}\bm{S}\bm{A} + \bm{N}_{y} = \bm{\tilde{S}}\bm{A} + \bm{N}_{y},
\end{aligned}
\end{equation}
where $\bm{S}\geq \bm{0}$ and $\bm{A}\geq \bm{0}$ denote a collection of pure spectral signatures (i.e., \textit{endmembers}) and corresponding fractional coefficients (i.e., \textit{abundances}) that meet the sum-to-one constraint (i.e., $\bm{1}^{\top}\bm{A}=\bm{1}^{\top}$), respectively, while $\bm{\tilde{S}}$ and $\bm{\tilde{A}}$ can be explained as the degraded $\bm{S}$ and $\bm{A}$ in spectral and spatial domains with the use of SRF and PSF. Thus, estimating $\bm{Z}$ is equivalent to finding $\bm{S}$ and $\bm{A}$, i.e., $\bm{Z}=\bm{S}\bm{A}$, by solving the problem (\ref{eq2}).

The model (\ref{eq2}) has been proven to be effective for the problem of mixed pixels in the HS-MS fusion \cite{yokoya2011coupled,lanaras2015hyperspectral,yokoya2017hyperspectral,dian2017hyperspectral,qu2018unsupervised,yao2020cross}. However, it seldom considers the high coupling between pixels in the same data source. For this reason, we learn the mappings ($f_x$ and $f_y$) for transforming $\bm{X}$ and $\bm{Y}$ to a latent image space (e.g., $\bm{\bar{X}}$, $\bm{\bar{Y}}$), where the pixel information can be easily separated between the same data source and better fused across different data sources. Therefore, the resulting model is 
\begin{equation}
\label{eq3}
\begin{aligned}
      &\bm{\bar{X}} = f_{x}(\bm{X})=f_{x}(\bm{S}\bm{\tilde{A}} + \bm{N}_{x}),\\
      &\bm{\bar{Y}} = f_{y}(\bm{Y})=f_{y}(\bm{\tilde{S}}\bm{A} + \bm{N}_{y}).
\end{aligned}
\end{equation}

In the following, we will build the DC-Net architecture step-by-step to perform the HS-MS fusion in the model (\ref{eq3}).

\subsection{Decoupled Network (D-Net)}
Owing to the powerful image-to-image transformation ability of GANs \cite{zhu2017unpaired,liu2017unsupervised,lee2018diverse}, we develop the D-Net to decouple the input images and recombine them (see Fig. \ref{fig:workflow}), making it possible for the recombined images to be fused more sufficiently. D-Net embeds the HS-MS images onto a common content space $\mathcal{C}$ and sensor-specific spaces, e.g., $\mathcal{S}_{\mathcal{X}}$ and $\mathcal{S}_{\mathcal{Y}}$. This process can be performed by cross-sensor (or domain-common) encoders $\{E_{\mathcal{X}}^{\mathcal{C}},E_{\mathcal{Y}}^{\mathcal{C}}\}$, sensor-specific encoders $\{E_{\mathcal{X}}^{\mathcal{S}},E_{\mathcal{Y}}^{\mathcal{S}}\}$, generators $\{G_{\mathcal{X}},G_{\mathcal{Y}}\}$, and a domain-common discriminator $D_{adv}^{\mathcal{C}}$. Given the input data ($\bm{X}$ and $\bm{Y}$), we then have the following transformations:
\begin{equation}
\label{eq4}
\begin{aligned}
      &\{\bm{X}^{\mathcal{C}},\bm{X}^{\mathcal{S}}\}=\{E_{\mathcal{X}}^{\mathcal{C}}(\bm{X}),E_{\mathcal{X}}^{\mathcal{S}}(\bm{X}_{\uparrow})\},\\
      &\{\bm{Y}^{\mathcal{C}},\bm{Y}^{\mathcal{S}}\}=\{E_{\mathcal{Y}}^{\mathcal{C}}(\bm{Y}),E_{\mathcal{Y}}^{\mathcal{S}}(\bm{Y}_{\downarrow})\},
\end{aligned}
\end{equation}
where $\bm{X}^{\mathcal{C}}\in \mathcal{C}$ and $\bm{Y}^{\mathcal{C}}\in \mathcal{C}$ are the common information between $\bm{X}$ and $\bm{Y}$, and $\bm{X}^{\mathcal{S}}\in \mathcal{S}_{\mathcal{X}}$ and $\bm{Y}^{\mathcal{S}}\in \mathcal{S}_{\mathcal{Y}}$ are the domain-specific information.

By interactively recombining the common and sensor-specific information, we can further obtain transformed images, i.e., $\mathcal{\bar{X}}$ and $\mathcal{\bar{Y}}$, via generators ($\{G_{\mathcal{X}},G_{\mathcal{Y}}\}$). This process can be formulated by $\bm{\bar{X}}=G_{\mathcal{X}}(\bm{X}^{\mathcal{C}},\bm{Y}^{\mathcal{S}})$ and $\bm{\bar{Y}} = G_{\mathcal{Y}}(\bm{Y}^{\mathcal{C}},\bm{X}^{\mathcal{S}})$. Such recombined images provide higher affinity between different sources (e.g., HS-MS), thereby yielding a greater potential for follow-up fusion.

\begin{figure}[!t]
	  \centering
			\includegraphics[width=0.25\textwidth]{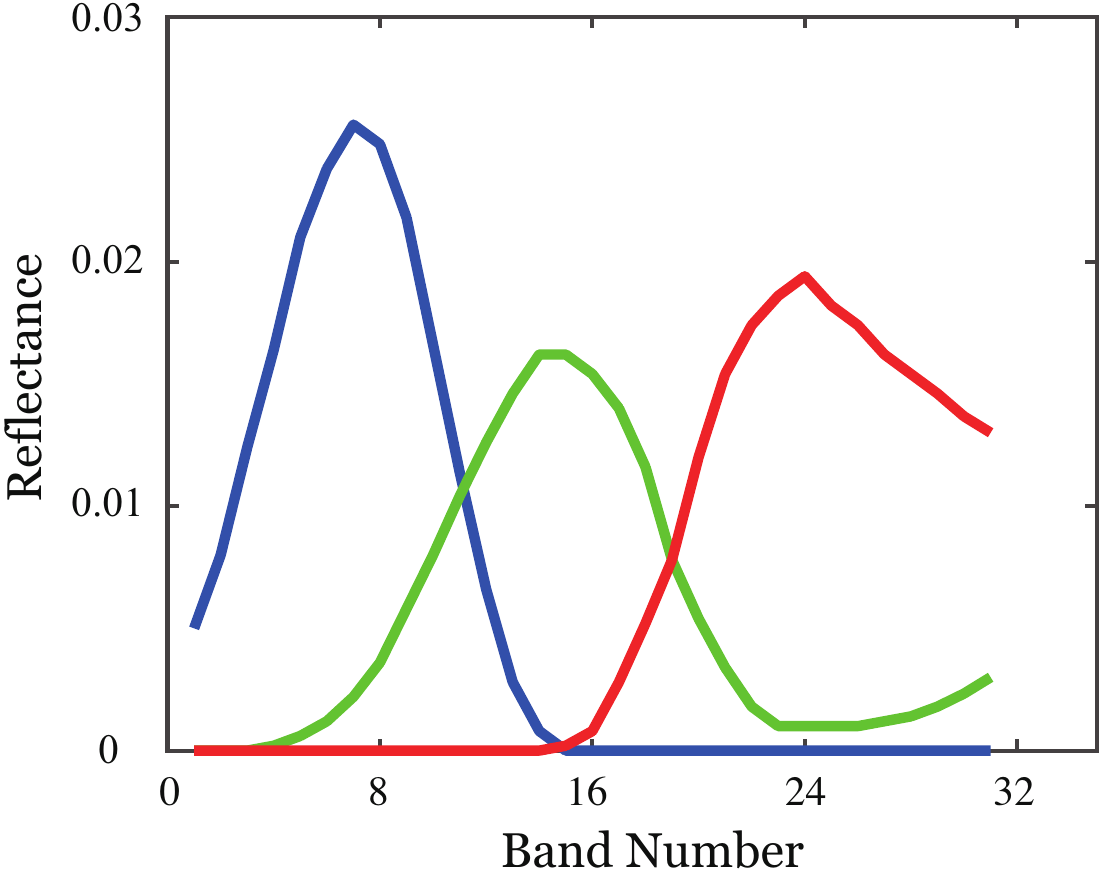}
        \caption{An example for SRF (i.e., $\bm{H}$) in RGB bands.}
        \label{fig:SRF}
\end{figure}

\subsection{Coupled Network (C-Net)}
Linking to the D-Net, C-Net unmixes the recombined HS and MS images into \textit{endmembers} ($\bm{S}$, $\bm{\tilde{S}}$) and \textit{abundances} ($\bm{\tilde{A}}$, $\bm{A}$). As shown in Fig.~\ref{fig:workflow}, C-Net is a coupled convolutional autoencoder, which consists of two encoders ($f_{u}$ and $g_{u}$) and two decoders ($f_{v}$ and $g_{v}$) for $\bm{\bar{X}}$ and $\bm{\bar{Y}}$, respectively. This can be formulated accordingly as
\begin{equation}
\label{eq6}
\begin{aligned}
      &\bm{\hat{X}}=\bm{S}\bm{\tilde{A}}=f(\bm{\bar{X}})=f_{v}(f_{u}(\bm{\bar{X}};\bm{W}_{f,u});\bm{W}_{f,v}),\\
      &\bm{\hat{Y}}=\bm{\tilde{S}}\bm{A}=g(\bm{\bar{Y}})=g_{v}(g_{u}(\bm{\bar{Y}};\bm{W}_{g,u});\bm{W}_{g,v}).
\end{aligned}
\end{equation}
$\bm{W}$ is the weights of C-Net, where $\bm{W}_{f,v}$ and $\bm{W}_{g,v}$ are specified to be a $1\times 1$ linear convolutional kernel, which can be well explained as \textit{endmembers}, i.e., $\bm{S}$ and $\bm{\tilde{S}}$. Note that we adopt the Clamp function \cite{zheng2020coupled} to meet the abundance non-negative constraint (ANC), while the abundance sum-to-one constraint (ASC), i.e.,$\bm{1}^{\top}\bm{A}=\bm{1}^{\top}$, can be guaranteed in the form of regularization. Using Eq. (\ref{eq6}), the to-be-estimated HrHS image $\bm{\hat{Z}}$ can be obtained by $\bm{\hat{Z}}=\bm{S}\bm{A}=f_{v}(g_{u}(\bm{\bar{Y}};\bm{W}_{g,u});\bm{W}_{f,v})$.

\vspace{3pt}
\noindent \textbf{Cycle consistency.} Coupled factors, i.e., SRF ($\bm{H}$) and PSF ($\bm{R}$), play a dominant role in the performance improvement of C-Net. Fig. \ref{fig:SRF} shows a SRF example w.r.t RGB bands. Unlike previous methods that assume the two functions to be known, we attempt to automatically learn them by designing a cycle consistency mechanism in networks, enabling it to be more applicable in reality. For simplicity, the cycle can be written as $(\bm{X},\bm{Y})\xrightarrow{\rm{D-Net}} (\bm{\bar{X}},\bm{\bar{Y}})\xrightarrow{\rm{C-Net}} \bm{\hat{Z}}\xrightarrow{(\bm{H},\bm{R})}(\bm{X},\bm{Y})$. More specifically, given the spectrum of the $i$-th pixel in $\bm{Z}$ (denoted as $\bm{z}_{i}$) and the corresponding spectrum of the $i$-th pixel and the $j$-th channel in $\bm{Y}$ (denoted as $y_{i,j}$), the degradation process can be then expressed as
\begin{equation}
\label{eq8}
\begin{aligned}
      y_{i,j}=\int_{\phi}\frac{\bm{h}_{j}(\eta)}{N_{h}}\bm{z}_{i}(\eta)d\eta,
\end{aligned}
\end{equation}
where $\phi$ denotes the support set w.r.t. the wavelength $\eta$, $\int\bm{h}_j(\eta)d\eta$ is the continuous representation of $\bm{H}$, and $N_{h}$ is a constant for scaled normalization. In our case, $\bm{H}$ can be approximated using a set of $1\times 1\times L$ convolutional kernels, e.g., $\bm{w}_{j}$, in a discretized way, i.e.,
\begin{equation}
\label{eq9}
\begin{aligned}
      y_{i,j}=\bm{z}_{j}\otimes \bm{w}_{j}=\sum\nolimits_\phi\frac{\bm{w}_{j}(\eta)}{N_{w}}\bm{z}_{i}(\eta).
\end{aligned}
\end{equation}
Similarly, spatial downsampling process performed by the PSF ($\bm{R}$) can be also simulated by means of convolutional operators with same scaling kernel and stride sizes.

As a result, the cycle consistency for both spatial and spectral degradation can be well organized in the C-Net as
\begin{equation}
\label{eq10}
\begin{aligned}
      \bm{X}=\bm{\hat{Z}}\bm{R},\;\; \bm{Y}=\bm{H}\bm{\hat{Z}},
\end{aligned}
\end{equation}
which is a good fit for the observation models in Eq. (\ref{eq1}) and helps us form a closed cycle chain.

\begin{figure}[!t]
	  \centering
			\includegraphics[width=0.46\textwidth]{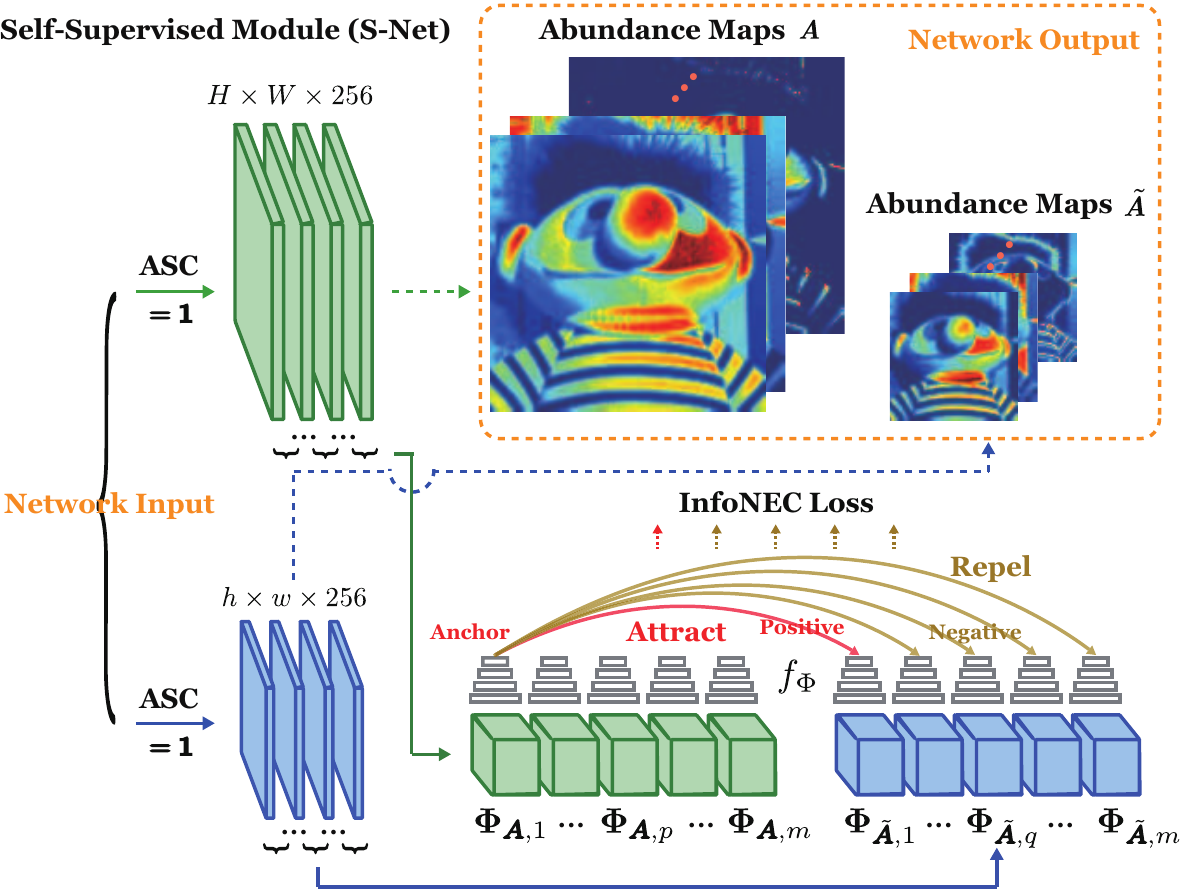}
        \caption{An illustration of S-Net module.}
\label{fig:S-Net}
\end{figure}

\subsection{Self-Supervised Subnetwork (S-Net)}
In C-Net, the spectral unmixing process for HS and MS images is relatively independent. This might lead to chaotic semantic correspondences on \textit{abundance maps} (or \textit{endmembers}) of HS-MS images, e.g., between $\bm{A}$ and $\bm{\tilde{A}}$ or $\bm{\tilde{S}}$ and $\bm{S}$. To this end, we design an effective self-supervised learning module (S-Net) in C-Net to obtain more plausible \textit{abundance maps} and \textit{endmembers}, further yielding a more accurate HS reconstruction result. Fig. \ref{fig:S-Net} illustrates the S-Net. 

S-Net aims at correcting the correspondence of \textit{abundance maps} of HS-MS images by learning high-level semantic alignment or consistency (in both order and category of materials), enabling a one-to-one match between $\bm{A}$ and $\bm{\tilde{A}}$. We call it as \textit{material or semantic consistency}. The \textit{abundance maps} of the same materials should be attracted to each other (as positive samples); otherwise, they are repelled (as negative samples). Considering the computational cost (since hundreds of abundance maps and spectral bundles usually need to be considered), we employ the same grouping strategy on $\bm{A}$ and $\bm{\tilde{A}}$ to make a group-to-group match. $\bm{A}$ and $\bm{\tilde{A}}$ can be grouped as
\begin{equation}
\label{eq11}
\begin{aligned}
      &\bm{A} = \lbrack \underbrace{\bm{A}_{1},\cdots}_{\bm{\Phi}_{\bm{A},1}},\cdots,\underbrace{\bm{A}_{k},\cdots}_{\bm{\Phi}_{\bm{A},p}},\cdots,\underbrace{\cdots,\bm{A}_{n}}_{\bm{\Phi}_{\bm{A},m}} \rbrack,\\
      &\bm{\tilde{A}} = \lbrack \underbrace{\bm{\tilde{A}}_{1},\cdots}_{\bm{\Phi}_{\bm{\tilde{A}},1}},\cdots,\underbrace{\bm{\tilde{A}}_{k},\cdots}_{\bm{\Phi}_{\bm{\tilde{A}},q}},\cdots,\underbrace{\cdots,\bm{\tilde{A}}_{n}}_{\bm{\Phi}_{\bm{\tilde{A}},m}} \rbrack,
\end{aligned}
\end{equation}
where $\bm{\Phi}_{\bm{A},p}$ and $\bm{\Phi}_{\bm{\tilde{A}},q}$ denote the grouping sets w.r.t. $\bm{A}$ and $\bm{\tilde{A}}$ for $\forall p,q\in\{1,\cdots,m\}$, respectively. We experimentally set the grouping parameter $m$ to be $8$. Furthermore, the representation vectors of $\bm{\Phi}_{\bm{A},p}$ and $\bm{\Phi}_{\bm{\tilde{A}},q}$, denoted as $\bm{z}$, can be obtained via a two-stream encoder or CNN, e.g., $f_{\bm{\Phi}_{\bm A}}$ and $f_{\bm{\Phi}_{\bm{\tilde{A}}}}$, which both consists of two blocks, i.e., [conv-relu-pool-fc]. The main difference lies in the pooling size and stride for $f_{\bm{\Phi}_{\bm A}}$ larger than those for $f_{\bm{\Phi}_{\bm{\tilde{A}}}}$ to guarantee the same size for final output features. We then have $\bm{z}_{\bm{A},p}=f_{\bm{\Phi}_{\bm A}}(\bm{\Phi}_{\bm{A},p})$ and $\bm{z}_{\bm{\tilde{A}},q}=f_{\bm{\Phi}_{\bm{\tilde{A}}}}(\bm{\Phi}_{\bm{\tilde{A}},q})$. In our S-Net, only when $p=q$, $\bm{z}_{\bm{\tilde{A}},q}$ is the positive sample of $\bm{z}_{\bm{A},p}$; otherwise ($p\neq q$), they are the negative sample pair.

\subsection{Network Training}
\noindent \textbf{Loss functions.} As shown in Fig. \ref{fig:workflow}, there are several important loss functions that need to be carefully considered.

In D-Net, the adversarial loss between the common content of $\bm{X}$ and $\bm{Y}$ can be expressed as
\begin{equation}
\label{eq12}
\begin{aligned}
      \mathcal{L}_{adv}=&\;\mathbb{E}_{x}[\frac{1}{2}\log D^{\mathcal{C}}(\bm{X}^{\mathcal{C}})+\frac{1}{2}\log(1-D^{\mathcal{C}}(\bm{X}^{\mathcal{C}})]\\
      +&\;\mathbb{E}_{y}[\frac{1}{2}\log D^{\mathcal{C}}(\bm{Y}_{\downarrow}^{\mathcal{C}})+\frac{1}{2}\log(1-D^{\mathcal{C}}(\bm{Y}_{\downarrow}^{\mathcal{C}})],
\end{aligned}
\end{equation}
where $\bm{X}^{\mathcal{C}}=E_{\mathcal{X}}^{\mathcal{C}}(\bm{X})$ and $\bm{Y}^{\mathcal{C}}=E_{\mathcal{Y}}^{\mathcal{C}}(\bm{Y})$.

The reconstruction loss stemmed from coupled convolutional autoencoders and the cycle consistency can be computed by means of Eqs. (\ref{eq6}) and (\ref{eq10}):
\begin{equation}
\label{eq13}
\begin{aligned}
      \mathcal{L}_{rec}&=\norm{\bm{\hat{X}}-\bm{X}}_{1}+\norm{\bm{\hat{Y}}-\bm{Y}}_{1}+\norm{\bm{\hat{X}}-\bm{Z}_{\downarrow,\bm{R}}}_{1}\\
      &+\norm{\bm{\hat{Y}}-\bm{Z}_{\downarrow,\bm{H}}}_{1} + \norm{\bm{\bar{X}}-\bm{X}}_{1} + \norm{\bm{\bar{Y}}-\bm{Y}}_{1},
\end{aligned}
\end{equation}
where the $\ell_{1}$-norm measure is applied to enhance the detailed perception in image reconstruction \cite{zhao2016loss}, and $\bm{\hat{Z}}_{\downarrow,\bm{R}}$ and $\bm{\hat{Z}}_{\downarrow,\bm{H}}$ denote $\bm{\hat{Z}}\bm{R}$ and $\bm{H}\bm{\hat{Z}}$, respectively.

Moreover, the ASC is satisfied using the following loss:
\begin{equation}
\label{eq14}
\begin{aligned}
      \mathcal{L}_{\rm ASC}=\norm{\bm{1}^{\top}-\bm{1}^{\top}\bm{A}}_{1} + \norm{\bm{1}^{\texorpdfstring{}{}}-\bm{1}^{\top}\bm{\tilde{A}}}_{1}.
\end{aligned}
\end{equation}
 
In S-Net, the InfoNCE loss \cite{oord2018representation} is optimized to measure the distances between positive and negative samples:
\begin{equation}
\label{eq15}
\begin{aligned}
      \mathcal{L}_{self}=-\mathbb{E}_{\bm{\Phi}}[\log \frac{\exp(f_{\bm{\Phi}}(\bm{\Phi}_{\bm{A},p})^{\top}f_{\bm{\Phi}}(\bm{\Phi}_{\bm{\tilde{A}},q}^{+}))}{\sum_{q=1}^{m}\exp(f_{\bm{\Phi}}(\bm{\Phi}_{\bm{A},p})^{\top}f_{\bm{\Phi}}(\bm{\Phi}_{\bm{\tilde{A}},q}))}],
\end{aligned}
\end{equation}
where $\bm{\Phi}_{\bm{A},p}$ and $\bm{\Phi}_{\bm{\tilde{A}},q}^{+}$ denote the anchor sample and the positive sample, respectively.

To sum up, our DC-Net is trained by minimizing the following overall loss parameterized by the set $\{\alpha,\beta,\gamma\}$:
\begin{equation}
\label{eq16}
\begin{aligned}
      \mathcal{L}=\mathcal{L}_{rec}+\alpha\mathcal{L}_{adv}+\beta\mathcal{L}_{\rm{ASC}}+\gamma\mathcal{L}_{self}.
\end{aligned}
\end{equation}

\noindent \textbf{Implementation details.} Our networks are implemented on the PyTorch platform, and the Adam \cite{kingma2014adam} optimizer is used to train the networks for 10000 epochs with a batch size of 1. The base learning rate starts with 0.005, which can be gradually updated by a linear decay schedule \cite{loshchilov2017decoupled}. We initialize all convolutional layers in our networks with Kaiming initialization \cite{he2015delving}. Furthermore, the hyper-parameters are experimentally determined using a grid search, and early stopping strategy is considered in the network training when the validation loss fails to decrease.

\begin{figure*}[!t]
	  \centering
			\includegraphics[width=0.9\textwidth]{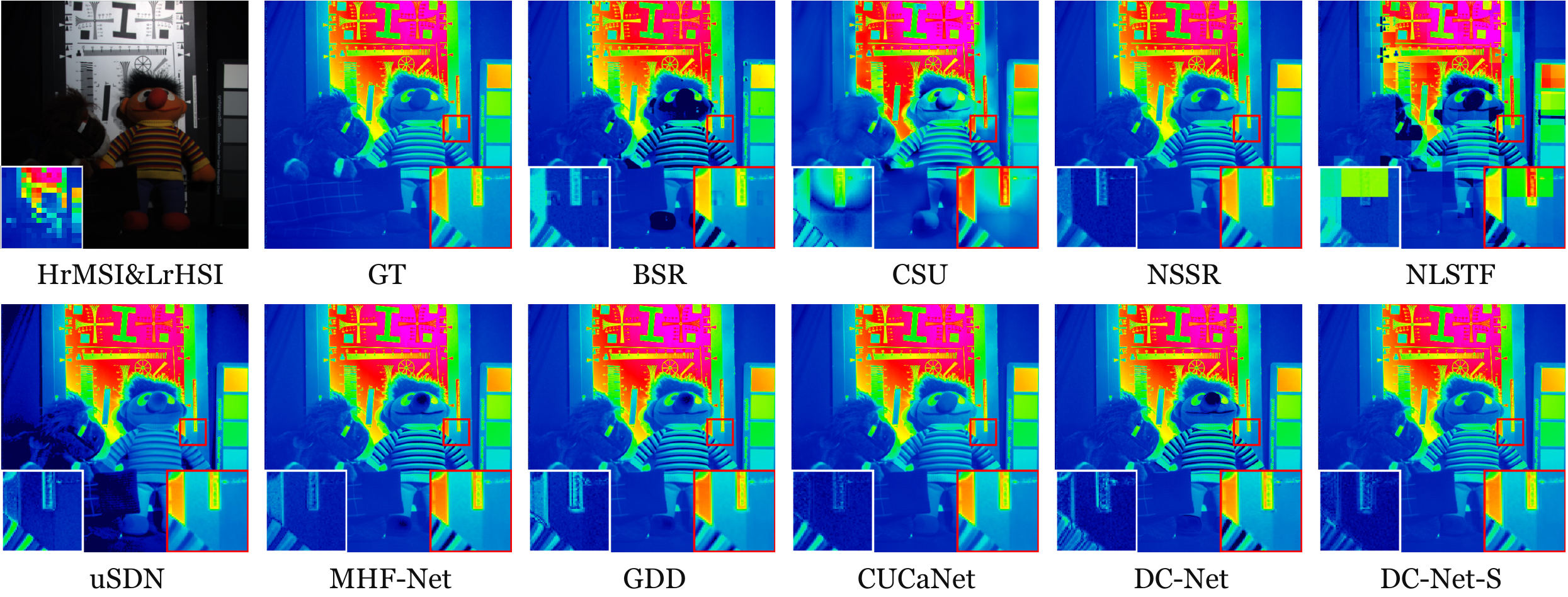}
        \caption{Visualization of super-resolved HS images obtained by all compared methods (CAVE: \textit{chart and stuffed toy}), where a ROI zoomed in $4$ times (bottom-right) and the difference image with GT (bottom-left) are highlighted.}
\label{fig:CAVE_Toy1}
\end{figure*}

\begin{figure*}[!t]
	  \centering
			\includegraphics[width=0.9\textwidth]{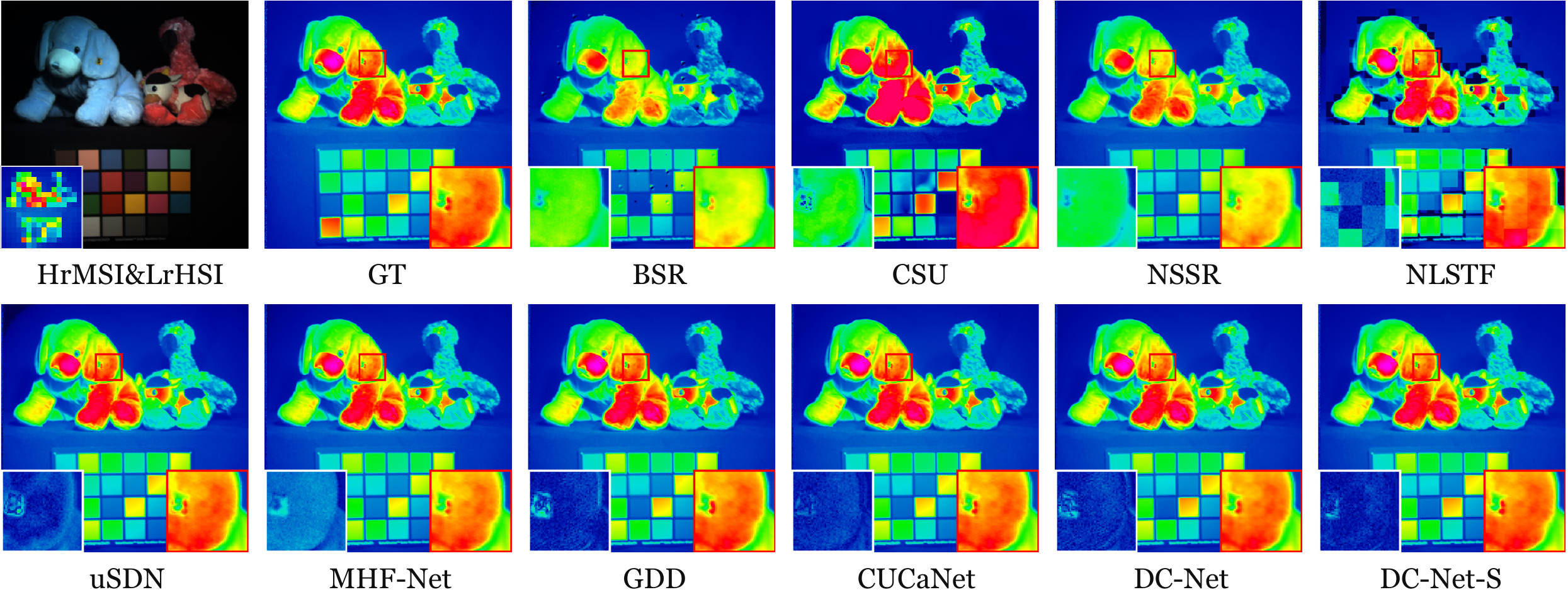}
        \caption{Visualization of super-resolved HS images obtained by all compared methods (CAVE: \textit{stuffed toys}), where a ROI zoomed in $4$ times (bottom-right) and the difference image with GT (bottom-left) are highlighted.}
\label{fig:CAVE_Toy2}
\end{figure*}

\section{Experimental Results}
In this section, we first review the datasets and experimental setting. Next, we verify our networks with extensive ablation studies. Last, the performance evaluation on simulated and real datasets is conducted to show the superiority of DC-Net in comparison with state-of-the-art methods.

\subsection{Dataset and Setup} 
Two public HS-MS datasets are used to evaluate the performance of DC-Net, i.e., CAVE dataset (indoor) \cite{yasuma2010generalized} and Chikusei dataset (remote sensing) \cite{yokoya2017hyperspectral}. 

The CAVE dataset consists of $32$ indoor HrHS images with the size of $512 \times 512\times 31$, covering the spectral wavelength from $400nm$ to $700nm$. The Chikusei dataset acquired by VNIR-C sensor comprises of $2517\times 2335$ pixels and $128$ spectral bands. We randomly crop the scene into $12$ non-overlapped subimages with spatial size of $576 \times 448$. Moreover, we divide the two datasets into training, validation, and testing sets in the proportions of 8:8:16 and 4:4:4, respectively, where training and validation sets are used to determine optimal hyperparameters.

For the two datasets, the HrMS images are generated by degrading the HrHS images in spectral domain with SRFs, while the LrHS images can be also simulated by spatially downsampling the HrHS images with the Wald's protocol \cite{Wald1997fusion} of a $32$ sampling ratio. In addition, the Nikon D700 sensor and Landsat-8 MS sensor \cite{barsi2014spectral} are selected as the SRFs for the CAVE and Chikusei datasets, respectively.

\vspace{3pt}
\noindent \textbf{Evaluation Metrics.} We evaluate the fusion performance quantitatively in terms of five widely-used picture quality indices (PQIs): peak signal-to-noise ratio (PSNR), spectral angle mapper (SAM) \cite{kruse1993spectral}, \textit{erreur relative globale adimensionnelle de synth{\`e}se} (ERGAS) \cite{wald2000quality}, structure similarity (SSIM) \cite{wang2004image}, and universal image quality index (UIQI) \cite{wang2002universal}.

\begin{table*}[!t]
    \centering
    \caption{Quantitative average performance of SOTA super-resolution methods over 16 testing images in terms of five different PQIs on CAVE dataset.}
    \resizebox{0.7\textwidth}{!}{
    \begin{tabular}{c|c||ccccc}
        \toprule[1.5pt]
        \multirow{2}{*}{Method} & \multirow{2}{*}{Ref.} & \multicolumn{5}{c}{Metric}\\
        \cline{3-7} & & PSNR & SAM & ERGAS & SSIM & UQI \\
        \hline\hline
        BSR & CVPR'15 \cite{akhtar2015bayesian} & 31.08 & 13.52 & 0.72 & 0.901 & 0.894 \\
        CSU & ICCV'15 \cite{lanaras2015hyperspectral} & 29.69 & 13.13 & 0.70 & 0.891 & 0.900 \\
        NSSR & TIP'16 \cite{dong2016hyperspectral} & 33.93 & 10.87 & 0.61 & 0.925 & 0.914 \\
        NLSTF & CVPR'17 \cite{dian2017hyperspectral} & 30.31 & 12.29 & 0.69 & 0.914 & 0.896 \\ 
        \hline
        uSDN & CVPR'18 \cite{qu2018unsupervised} & 35.60 & 9.97 & 0.53 & 0.934 & 0.921 \\
        MHF-Net & CVPR'19 \cite{xie2019multispectral} & 37.41 & 7.33 & 0.45 & 0.965 & 0.958 \\
        GDD & ECCV'20 \cite{uezato2020guided} & 37.06 & 6.71 & 0.40 & 0.971 & 0.960 \\
        CUCaNet & ECCV'20 \cite{yao2020cross} & 37.29 & 7.03 & 0.46 & 0.960 & 0.951 \\
        \hline
        DC-Net & -- & 38.12 & 6.48 & 0.39 & 0.973 & 0.977 \\
        DC-Net-S & -- & \bf 38.68 & \bf 6.15 & \bf 0.34 & \bf 0.978 & \bf 0.980\\
        \hline\hline
        Ideal Value & -- & $\uparrow$ & $\downarrow \bm{0}$ & $\downarrow \bm{0}$ & $\uparrow \bm{1}$ & $\uparrow \bm{1}$\\ 
        \bottomrule[1.5pt]
    \end{tabular}
    }
    \label{tab:CAVE}
\end{table*}

\begin{figure*}[!t]
	  \centering
			\includegraphics[width=1\textwidth]{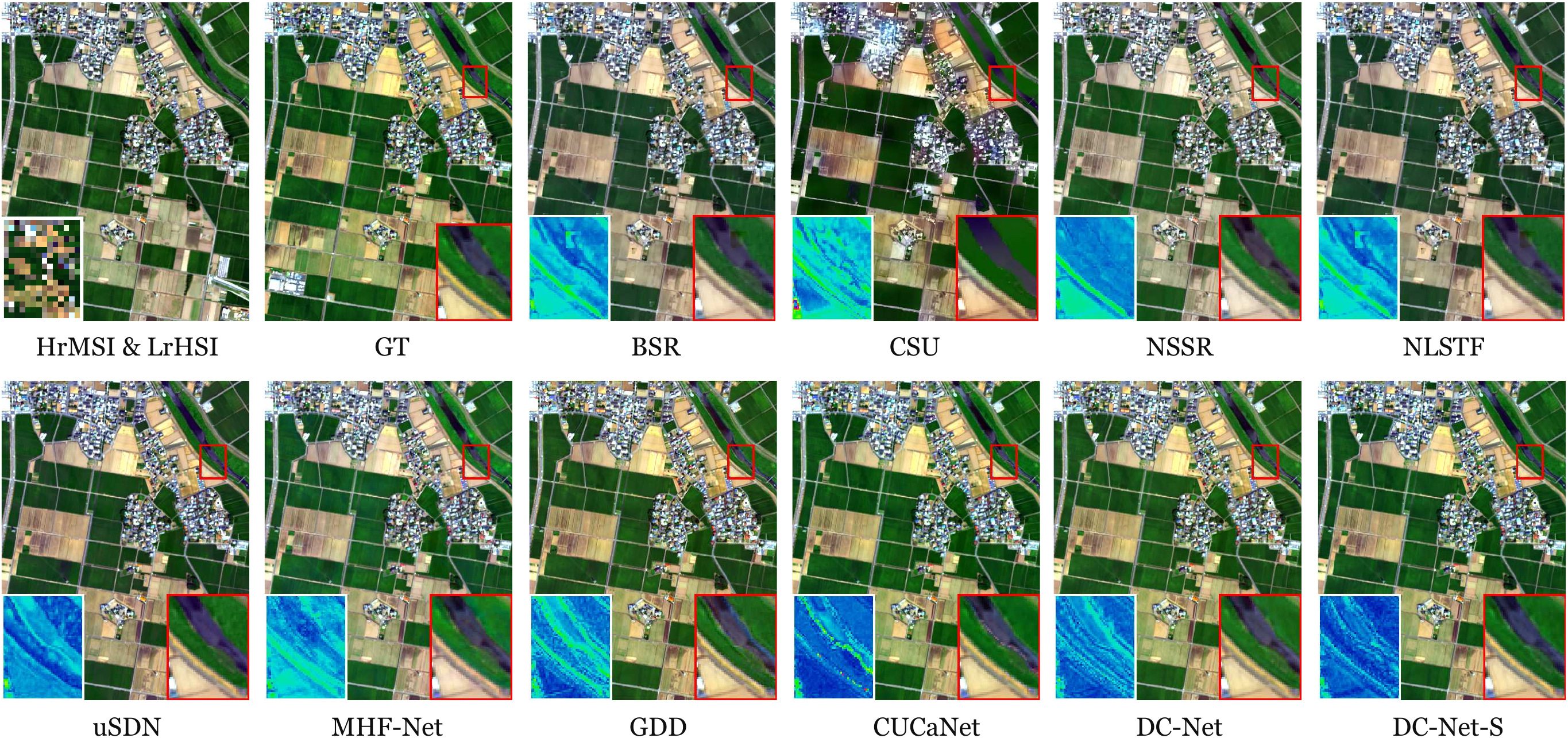}
        \caption{Visualization (R: 63, G: 35, B: 11) of super-resolved HS images obtained by all compared methods on Chikusei data, where a ROI zoomed in $4$ times and the difference image with GT are highlighted for more detailed observation.}
\label{fig:Chikusei}
\end{figure*}

\subsection{Comparison with State-of-the-arts}
We evaluate the performance of DC-Net both quantitatively and qualitatively in comparison with state-of-the-art (SOTA) models in the HS-SR task, including traditional SOTA methods: BSR \cite{akhtar2015bayesian}\footnote{\scriptsize{\url{https://github.com/junjun-jiang/Hyperspectral-Image-Super-Resolution-Benchmark}}}, CSU \cite{lanaras2015hyperspectral}\footnote{\scriptsize{\url{https:/github.com/lanha/SupResPALM}}}, NSSR \cite{dong2016hyperspectral}\footnote{\scriptsize{\url{http://see.xidian.edu.cn/faculty/wsdong}}}, NLSTF \cite{dian2017hyperspectral}\footnote{\scriptsize{\url{https://github.com/renweidian/NLSTF}}}, and DL-based SOTA methods: uSDN \cite{qu2018unsupervised}\footnote{\scriptsize{\url{https://github.com/aicip/uSDN}}}, MHF-Net \cite{xie2019multispectral}\footnote{\scriptsize{\url{https://github.com/XieQi2015/MHF-net}}}, GDD \cite{uezato2020guided}\footnote{\scriptsize{\url{https://github.com/tuezato/guided-deep-decoder}}}, CUCaNet \cite{yao2020cross}\footnote{\scriptsize{\url{https://github.com/danfenghong/ECCV2020\_CUCaNet}}}, as well as the DC-Net without S-Net module. We maintain the same experimental configurations for all compared methods as in their original literature as much as possible.

\subsubsection{Experiments on CAVE Data} Table~\ref{tab:CAVE} lists the average results over 16 testing images on CAVE dataset to make quantitative comparison between all compared SOTA methods and the DC-Net and DC-Net-S in terms of five PQIs.

\begin{figure*}[!t]
	  \centering
			\includegraphics[width=1\textwidth]{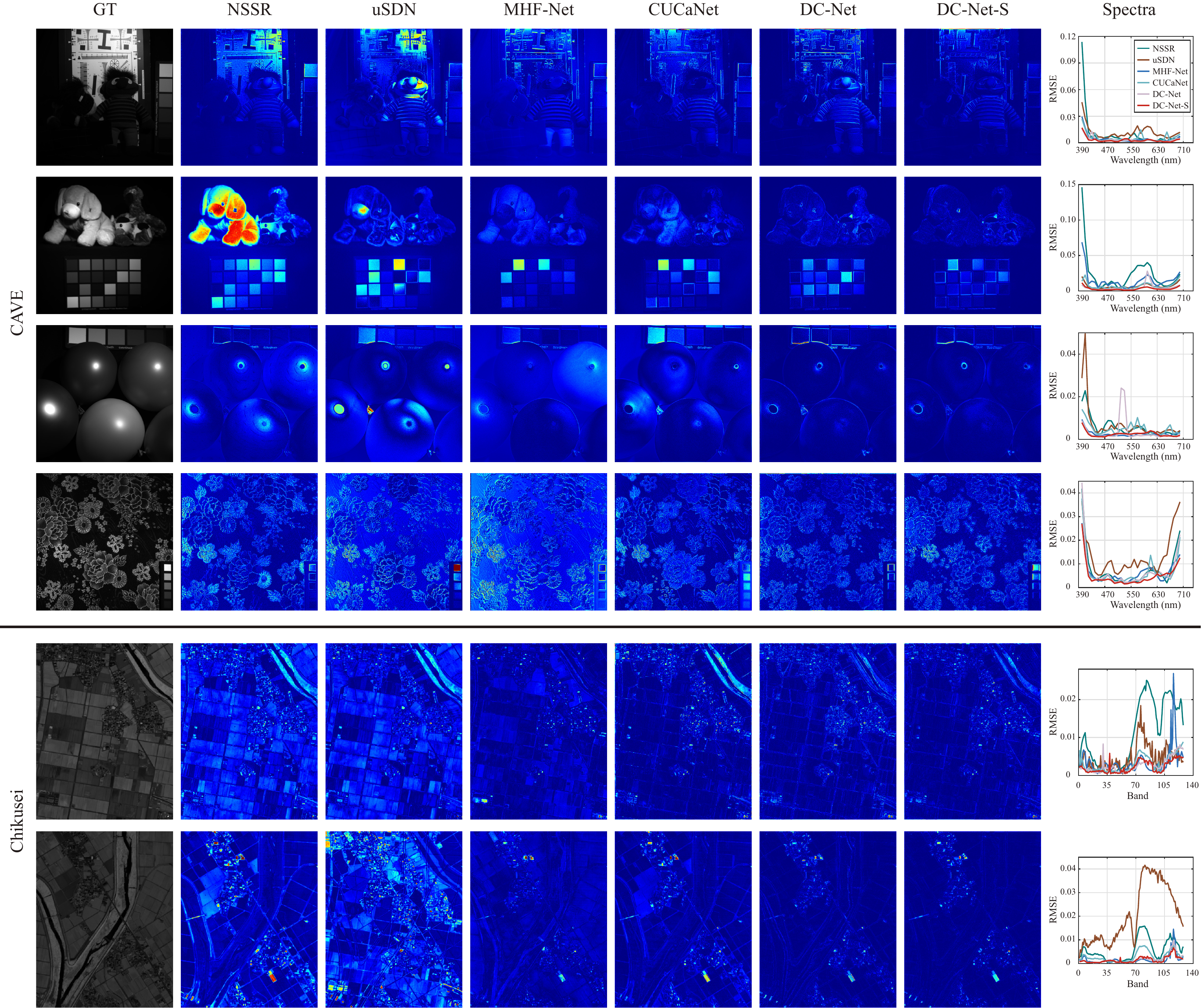}
        \caption{Visual evaluation on CAVE and Chikusei datasets, respectively. The GT, the residual maps between GT maps and super-solved HS results obtained by compared methods (e.g., NSSR, uSDN, MHF-Net, CUCaNet) and our proposed DC-Net and DC-Net-S. The RMSE values of HS images for these methods along the spectra (covering the wavelength) are also given corresponding to the six studied scenes.}
\label{fig:RMSE-CAVE}
\end{figure*}

Overall, BSR, as a pioneer work in HS-SR, reconstructs the high-quality HS image from the Bayesian statistic point of view and achieves competitive performance. Despite relatively lower PSNR and SSIM (\textit{cf.} BSR), CSU can obtain better POIs in SAM, ERGAS, and UQI, owing in large part to the use of coupled spectral unmixing mechanism. Beyond the CSU, NLSTF fuses HS and MS images in a coupled tensor form with non-local spatial information preservation, yielding better performance in all indices except the UQI. By virtue of non-negativity and structured modeling, NSSR outperforms other traditional SOTA methods in HrHS image reconstruction. As an early-stage DL-based method, the fusion performance of uSDN is moderately inferior to that of other DL-based competitors, but is clearly higher than that of those non-DL algorithms. It is more noteworthy that the results obtained by unsupervised GDD and CUCaNet are basically comparable to that of the supervised MHF-Net (even better in certain indices). DC-Net behaves superiorly compared to existing SOTA DL approaches, implying that the decoupled-and-coupled strategy creates the ``one plus one greater than two'' effect. Further, S-Net module can learn high-level semantic consistency from abundance maps and embed interpretable and physically meaningful information into the DC-Net, enabling the best performance (i.e., DC-Net-S).

Figs.~\ref{fig:CAVE_Toy1} and \ref{fig:CAVE_Toy2} visualize the super-resolved HS image in the $11$-th band (\textit{ca.} $500nm$) for \textit{chart and stuffed toy}. Further, two regions of interest (ROIs) are selected to highlight the visually detailed differences of all compared methods. By and large, DL-based models outperform evidently compared to traditional methods, particularly in coarser-grained structure of objects. The results of the proposed methods are superior to other unsupervised ones (e.g., uSDN, GDD, CUCaNet), while DC-Net-S performs better than DC-Net in terms of detailed appearances. Despite similar visual results (\textit{cf.} MHF-Net), our networks, either DC-Net or DC-Net-S, are capable of better recovering textural details and color cues that approximate to ground truth (GT).

\begin{table*}[!t]
    \centering
    \caption{Quantitative performance comparison between the same SOTA super-resolution methods in terms of five different PQIs on Chikusei dataset.}
     \resizebox{0.7\textwidth}{!}{
    \begin{tabular}{c|c||ccccc}
        \toprule[1.5pt]
        \multirow{2}{*}{Method} & \multirow{2}{*}{Ref.} & \multicolumn{5}{c}{Metric}\\
        \cline{3-7} & & PSNR & SAM & ERGAS & SSIM & UQI \\
        \hline\hline
        BSR & CVPR'15 \cite{akhtar2015bayesian} & 38.13 & 4.54 & 0.60 & 0.948 & 0.963 \\
        CSU & ICCV'15 \cite{lanaras2015hyperspectral} & 36.96 & 5.20 & 0.73 & 0.898 & 0.937 \\
        NSSR & TIP'16 \cite{dong2016hyperspectral} & 42.30 & 3.12 & 0.48 & 0.970 & 0.984 \\
        NLSTF & CVPR'17 \cite{dian2017hyperspectral} & 37.04 & 6.11 & 0.70 & 0.901 & 0.925 \\ 
        \hline
        uSDN & CVPR'18 \cite{qu2018unsupervised} & 40.02 & 2.96 & 0.41 & 0.964 & 0.981 \\
        MHF-Net & CVPR'19 \cite{xie2019multispectral} & 42.43 & 3.01 & 0.44 & 0.977 & 0.982 \\
        GDD & ECCV'20 \cite{uezato2020guided} & 41.58 & 3.04 & 0.43 & 0.974 & 0.984\\
        CUCaNet & ECCV'20 \cite{yao2020cross} & 41.79 & 2.84 & 0.40 & 0.975 & 0.988 \\
        \hline
        DC-Net & -- & 42.08 & 2.78 & 0.38 & 0.977 & 0.989 \\
        DC-Net-S & -- & \bf 42.56 & \bf 2.55 & \bf 0.34 & \bf 0.979 & \bf 0.992 \\
        \hline\hline
        Ideal Value & -- & $\uparrow$ & $\downarrow \bm{0}$ & $\downarrow \bm{0}$ & $\uparrow \bm{1}$ & $\uparrow \bm{1}$\\
        \bottomrule[1.5pt]
    \end{tabular}}
    \label{tab:Chikusei}
\end{table*}

\begin{figure}[!t]
	  \centering
			\includegraphics[width=0.48\textwidth]{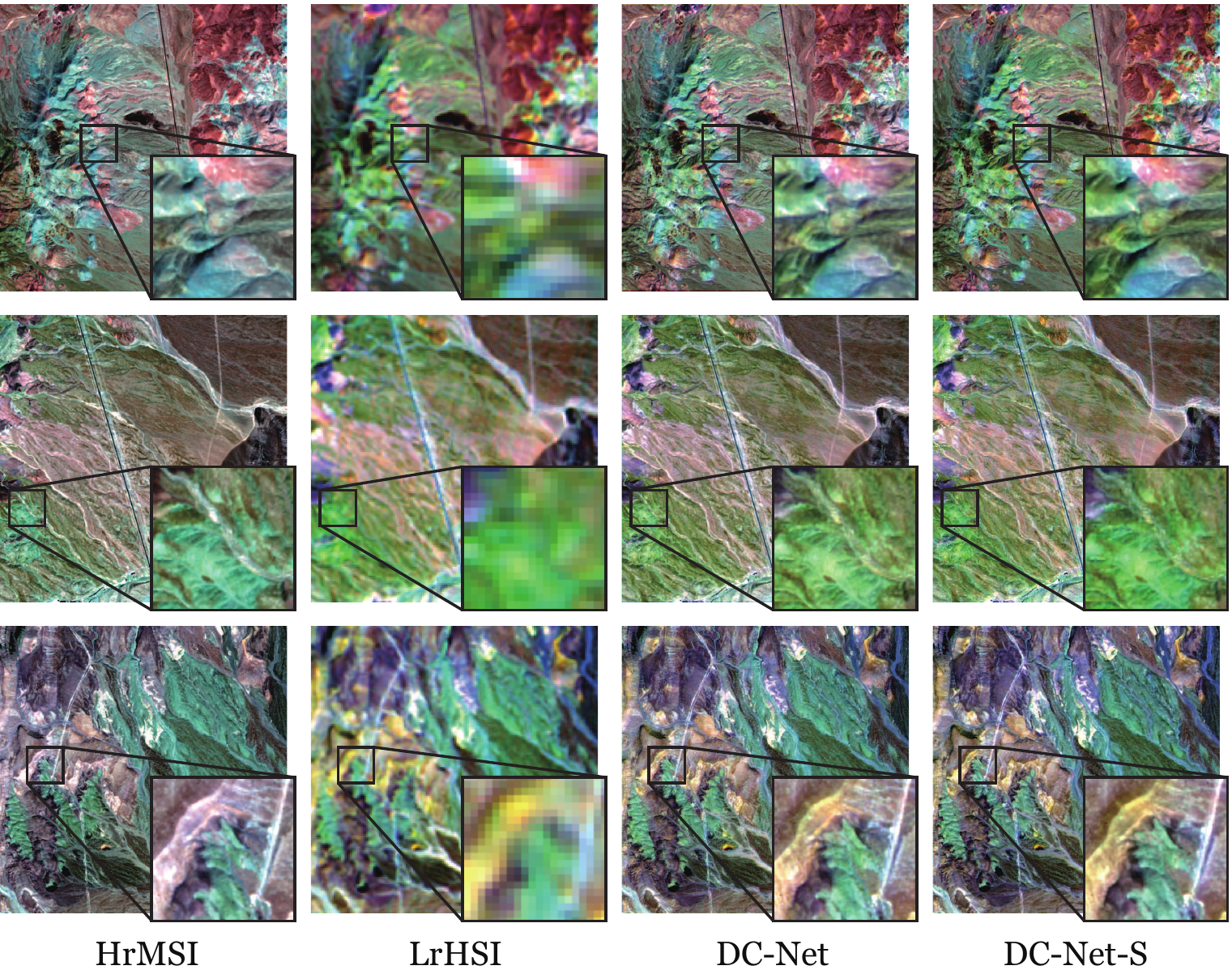}
        \caption{Visualization of super-resolved HrHS fused by WV-3 HrMS (image courtesy Maxar) and Hyperion LrHS using our method (real data of Cuprite: three sub-regions).}
\label{fig:Cuprite}
\end{figure}

\subsubsection{Experiments on Chikusei Data} Quantitative performance comparison on Chikusei dataset is given in Table~\ref{tab:Chikusei}, where there is a basically identical trend with that on CAVE dataset (see Table~\ref{tab:CAVE}). Remote sensing images (due to the lower spatial resolution) are not as complicated as indoor images, all these methods, therefore, tend to super-resolve HS images with higher quality on Chikusei dataset (\textit{cf.} CAVE). Significantly, the performance of our DC-Net-S is still superior to other competitors w.r.t. all PQIs, demonstrating its superiority in the HS-SR task. However, the most noteworthy point lies in that MHF-Net obtains competitive results (fully comparable to DC-Net without S-Net module), since its accuracy relies greatly on supervised training on plenty of sample pairs. Fig.~\ref{fig:Chikusei} gives a scene example and shows the composite HrHS images obtained by all competing methods with a ROI zoomed in $4$ times for better observing visual differences. 

\begin{table*}[!t]
    \centering
    \caption{Ablation analysis of the proposed DC-Net with a combination of different modules on the CAVE dataset.}
    \resizebox{0.6\textwidth}{!}{
    \begin{tabular}{c|ccccc||cc}
        \toprule[1.5pt]
        \multirow{2}{*}{Method} & \multicolumn{5}{c||}{Module} & \multicolumn{2}{c}{Metric} \\
        \cline{2-8} & ANC & ASC & D-Net & C-Net & S-Net & PSNR & SAM \\
        \hline\hline
        C-Net & \xmark & \xmark & \xmark & \cmark & \xmark & 32.56 & 14.80\\
        C-Net & \cmark & \xmark & \xmark & \cmark & \xmark & 36.69 & 7.43\\
        C-Net & \cmark & \cmark & \xmark & \cmark & \xmark & 36.97 & 6.84\\
        DC-Net & \cmark & \cmark &\cmark & \cmark & \xmark & 38.36 & 4.77\\ 
        DC-Net-S & \cmark & \cmark & \cmark & \cmark & \cmark & \bf 39.15 & \bf 4.51\\
        \hline\hline
        Ideal Value & -- & -- & -- & -- & -- & $\uparrow$ & $\downarrow \bm{0}$\\ 
        \bottomrule[1.5pt]
    \end{tabular}}
    \label{tab:ablation}
\end{table*}

\subsubsection{Visual Quality Evaluation}
To evaluate the visual quality of HS super-solved results more finely, we select several representative images from CAVE (four scenes) and Chikusei (two scenes) datasets, respectively, and show the corresponding residual maps between GT maps and super-solved HS results obtained by different compared methods, e.g., NSSR, uSDN, MHF-Net, CUCaNet, and our proposed DC-Net and DC-Net-S, in Fig. \ref{fig:RMSE-CAVE}. Furthermore, the RMSE values of HS images in each band are shown along spectra corresponding to the aforementioned six scenes. Intuitively, our methods perform better than other competitors from the perspective of residual maps, which demonstrates that our methods can provide finer spatial details. More importantly, the stability and generalization ability of our DC-Nets are higher than compared networks in different scenes, showing the effectiveness to a great extent. The band-wise RMES results obtained by our models are obviously similar and closer to the GT in the six investigated scenes. This, to some extent, indicates that the sequentially spectral information can be well recovered and retained by our approaches.

\subsection{Evaluation on Real Data}
To evaluate the performance of our DC-Net effectively, we collect a set of real and registered MS and HS scenes captured by WorldView-3 and Hyperion satellite missions, respectively. Three sub-regions are selected to visualize the quality comparison of composite HrHS products, as shown in Fig.~\ref{fig:Cuprite}. Each sub-region consists of HrMS and LrHS images with the size of $640\times 640$ pixels with $8$ bands at a $7.5m$ ground sampling distance (GSD) and $160\times 160$ pixels with $167$ spectral channels at a GSD of $30m$, respectively, while the real HrHS image is not available. In Fig.~\ref{fig:Cuprite}, both DC-Net and DC-Net-S can restore high resolution HrHS results. Despite clear results obtained by the two methods, the latter gives a higher degree of visual verisimilitude. Compared to the DC-Net, the structural and textural appearances of the DC-Net-S sit at the same level as the HrMS image and even beyond, while its color and brightness are closer to that of the LrHS image, showing the superiority of the proposed methods in the HS-SR task.

\subsection{Model Analysis}
\subsubsection{Ablation Studies} We investigate the performance gain by stepwise adding different modules (or subnetworks, i.e., D-Net and S-Net) in networks. We also study the importance of two physical constraints (ANC and ASC) related to spectral unmixing model and their effects on the quality of HS-SR. To verify the effectiveness of the ablation analysis, we report the average results in terms of PSNR and SAM indices over 8 images on the validation set of the CAVE dataset by maximizing the stepwise performance using optimal parameters, as listed in Table \ref{tab:ablation}.

In detail, C-Net without ANC and ASC yields relatively poor performance, which to some extent indicates the importance of the two important physical constraints in the spectral unmixing-inspired HS-SR network. By turning on the ANC, the fusion results of C-Net increase dramatically (beyond $4$ PNSR value and around halved SAM value), What is better still, the joint use of ANC and ASC can further bring the performance improvement in both PSNR and SAM compared to ANC used alone in C-Net. As expected, DC-Net after plugging the D-Net performs observably better than C-Net at an increase of about $1.5$ PSNR value and a decrease of just over $2$ SAM value. More remarkably, the self-supervised module (S-Net) is capable of breaking through the existing ``bottleneck'' in DC-Net by enforcing the semantic consistency on abundance maps ($\bm{A}$ and $\bm{\tilde{A}}$).

\begin{table}[!t]
    \centering
    \caption{Parameter size and running time of our DC-Net-S.}
    \resizebox{0.4\textwidth}{!}{
    \begin{tabular}{c|c|ccc}
    \toprule[1.5pt]
    \multirow{2}{*}{Model} & \multirow{2}{*}{Size (MB)} & \multicolumn{3}{c}{Average Run Time (min.)} \\
    \cline{3-5}
    & & CAVE & Chikusei & Cuprite\\
    \hline\hline
    DC-Net-S & 3.43 & 101 & 312 & 156 \\
    \bottomrule[1.5pt]
    \end{tabular}}
\label{table:runtime}
\end{table}

\subsubsection{Computational Analysis} 
We list the size of network parameters (i.e., DC-Net-S) and the inference time per image on a PC with one NVIDIA GeForce GTX 1080Ti GPU in Table. \ref{table:runtime}. Our network parameters only have 3.43MB, and since our method is unsupervised, the inference time is also desirable and acceptable in such a PC setting.

\section{Conclusion}
In this paper, we propose a novel subpixel-level HS-SR framework, i.e., DC-Net, by fully considering the effects of affinity of the same data sources and exclusivity across HS-MS data and mixed pixels within one pixel. Inspired by the spectral unmixing, DC-Net is capable of utilizing the intrinsic properties of HS-MS images effectively for the fusion task, which is more applicable to various real cases. We further optimize the network performance via devising an effective self-supervised module. Extensive experiments conducted on simulated and real data show the superiority of DC-Net in HS-SR over current SOTA fusion methods. We found, however, that there exists spectral degradation in the HS super-resolution products of the real data. As a result, we will pay more attention to the development of the new module or the use of the new network architecture to preserve the spectral sequentiality as much as possible in the process of improving the resolution.



\bibliographystyle{ieeetr}
\bibliography{HDF_ref}

\end{document}